# A Statistical Analysis of the SOT-Hinode Observations of Solar Spicules and their Wave-like Behavior

E. Tavabi[1, 2, 3], S. Koutchmy[2, 3] and A. Ajabshirizadeh[3, 4]

1-Payame Noor University of Zanjan, Zanjan, Iran

2- Institut d'Astrophysique de Paris (CNRS) & UPMC, UMR 7095, 98 Bis Boulevard Arago, F-75014 Paris, France,

3- Research Institute for Astronomy & Astrophysics of Maragha (RIAAM),

4- Department of Theoretical Physics and Astrophysics, Tabriz University, 51664 Tabriz, Iran.

(E-mail: tavabi@iap.fr, koutchmy@iap.fr, a-adjab@tabrizu.ac.ir).

## Abstract

Spicules are an important very dynamical and rather cool structure extending between the solar surface and the corona. They are partly filling the space inside the chromosphere and they are surrounded by a transition thin layer. New space observations taken with the SOT of the Hinode mission shed some light on their still mysterious formation and dynamics. Here we restrict the analysis to the most radial and the most interesting polar spicules situated at the base of the fast solar wind of coronal holes.

We consider a first important parameter of spicules as observed above the solar visible limb: their apparent diameter as a function of the height above the limb which determines their aspect ratio and leads to the discussion of their magnetic origin using the flux tube approximation. We found that indeed spicules show a whole range of diameters, including unresolved "interacting spicules" (I-S), depending of the definition chosen to characterize this ubiquitous dynamical phenomenon occurring into a low coronal surrounding. Superposition effects along the line of sight have to be taken into account in order to correctly measure individual spicules and look at I-S. We take advantage of the so-called Madmax operator to reduce these effects and improve the visibility of these hair-like features. An excellent time sequence of images obtained above a polar region with the Hinode SOT through the HCaII filter with a cadence of 8 s was selected for analysis. 1-D Fourier amplitude spectra (AS) made at different heights above the limb are shown for the first time. A definite signature in the 0.18 to 0.25 Mm range exists, corresponding to the occurrence of the newly discovered type II spicules and, even more impressively, large Fourier amplitudes are observed in the 0.3 to the 1.2 Mm range of diameters and spacing, in rough agreement with what historical works were reporting. Additionally, some statistically significant behavior, based on AS computed for different heights above the limb, is discussed.
“Time slice or x-t diagrams” revealing the dynamical behavior of spicules are also analyzed. They show that most of spicules have multiple structures (similarly to the doublet spicules) and they show impressive transverse periodic fluctuations which were interpreted as upward kink or Alfven waves. Evidence of the helical motion in spicules is now well evidenced, the typical periods of the apparent oscillation being around 120 sec. A fine analysis of the time-slice diagram as a function of the effective heights shows an interesting new feature near the 2 Mm height. We speculate on the interpretation of this feature as being a result of the dynamical specificities of the spicule helical motion as seen in these unprecedented high resolution HCaII line emission time series.

## 1- Introduction

The region near the interface between the solar photosphere, with a decreasing upward radial temperature reaching a minimum value near 4500 degrees, and the surrounding extended 1 million degrees corona, is the subject of intense investigations in order to understand what is the heating mechanism and what is the mechanism of supplying the ionized plasma to upward layers and, ultimately, to the solar wind. It is a region where the magnetic field is rapidly filling the whole space around and where magnetic forces become dominant, making the approximation of the hydrostatic equilibrium rather inappropriate. And the region is showing highly dynamical phenomena at small scale.

Inside this region, the ubiquitous and extremely prolific spicules are spiky or jet-like structures and they were seen for a long time in emission of low-excitation chromospheric lines when observed off limb of the Sun. They ascend with an average apparent velocity of about 25 km/s and their line profile changes drastically as a function of the height above the visible limb, see Michard (1954), Athay and Bessey (1964) and Giovanelli, Michard and Mouradian (1965) which partly reflects the change of their behavior and the behavior of the inter-spicular atmosphere. It is clear that spicules are surrounded by a hot coronal atmosphere which penetrates down to the 2 Mm heights and, accordingly, an extremely thin transition region surrounds each spicule, making it visible in EUV emission lines like the HeII and the CIV lines, see Bohlin et al. (1975), Cook et al, (1983). .

From the study of filtergrams, spicules reach a maximum height of 9 Mm, and then either fade out or fall back with a velocity similar to ascent velocity (Beckers 1972). The temperature in a spicule is almost constant along its length (about 10,000-15,000 K from the line profile analysis), while the density in it is about ten to a hundred times greater than that in an inter-spicular region, see the historical exhaustive reviews of Beckers (1968- 1972). Their bases or feet lie presumably near 1 to 2 Mm above the photosphere, their root being disconnected from the solar surface or invisible. They are located at the edge of the supergranular cells where the chromospheric network is located, forming a kind of rosette. They are estimated to cover about or less than 1% of the solar surface. Their typical electron density is measured to be approximately $10^{11}$ $cm^{-3}$ for a mature average spicule (without taking into account a possible non-homogeneity across the flux tube which confines it). From the classical Beckers's review, measurements of spicule diameters range from 0.15-2.5 Mm (Beckers 1968-72), including a visual early historical measurement by A. Secchi (1877) giving 0.15 Mm. The observed diameter of spicules is indeed determined by the telescope diffraction limit and by the atmospheric seeing when long exposure times are used. Most of publications gives only estimates of the diameter from images which are, therefore, rather unreliable, ***but*** all authors discusses this point and most of them tried to correct their data from these effects. A good example of the correction introduced to deduce the diameters of spicules observed on filtergrams is given in the work of Dunn (1960) and later, of Lynch, Dunn and Beckers (1973).

In order to get the best "historical" value to further select a reliable method to be used to enhanced their visibility on new observations, we briefly re-examined the most recognized papers on the subject, including recent papers. We give in Table 1 a list of all estimations of spicule diameters that were found. Let us make some comments on these results. The most reliable data in this list are undoubtedly the microphotometer measurements by R.B. Dunn (he measured spicule diameters from the intensity tracings using many circular microphotometer scans). Each spicule corresponds to a local intensity increase in an otherwise rather amorphous intensity background resulting from scattered light, unresolved spicule, and perhaps some interspicular emission, both because of the excellent quality of the negatives used and of the instrumental design, the spicules diameter being defined as the full width at half intensity (FWHM). This diameter probably varies from spicule to spicule, with most of the spicules (>90%) having diameters between 0.4 and 1.5 Mm, from Dunn (1960) and a real dispersion is established. In addition, from the excellent Dunn's movies, Mouradian (1967) has reported a strong change of spicule diameter with time and, accordingly, with the heights. The phenomenon was called a "diffusion" of spicules into the surrounding inter-spicular atmosphere but we will not consider it in detail here. The velocity of expansion is about 22 km/s. This expansion begins after the spicule has reached a maximum local intensity and lasts for at least 1.5 min. A similar behavior was reported by Koutchmy and Loucif (1991) for "small" spicules observed using the 40' coronagraph at Sacramento Peak Observatory. New observations report that almost all spicules have an averaged diameter of ~ 0.5 Mm (Foukal 1990; Sterling 1998) although reported diameters of about 0.7 Mm (1″) and less still could depend of the resolution limit of the telescope, in the opinion of De Pontieu et al. (2007a) who also reported in De Pontieu et al. (2007b) the observations of a new type of spicules of shorter lifetimes and even smaller size that they call type II spicules. According to the Table 1, no report exists of diameter above 2.5 Mm and the most probable value (not the smallest value!) is near or under 1 Mm. Of course these values are still affected by the image quality and the confusion produced by the overlapping effects (or superposition of images of apparently close spicules due to line of sight effect) such that further, we will use a suitable image processing to significantly reduce this effect in order to analyze their behavior, without effecting their fundamental properties as defined by their reported diameters of Table 1.

TABLE I – Reported spicule diameters

| Investigators | Year | Spectral range | Average Diameter (Mm) | Comments | Method |
|---|---|---|---|---|---|
| Secchi | 1877 | H$\alpha$ | 0.15 | Visual | Visual |
| Menzel | 1931 | Ca H&K | 1..8 | total eclipse | Estimate |
| Roberts | 1945 | H$\alpha$ | 2..5 | Coronograph | Estimate |
| Mohler | 1951 | Ca H&K | 1..3 | total eclipse | Estimate |
| Rush & Roberts | 1954 | H$\alpha$ | 2..0 | Telescope | Estimate |
| Woltjer | 1954 | H$\alpha$ | 2..5 | Photometric | Microphotometer |
| Athay | 1959 | H$\alpha$ | 0.7 | Telescope | Image |
| Dunn | 1960 | H$\alpha$ | 0.81 $\in [0.2, 0.95]$ | Refractor | Microphotometer |
| Krat & Krat | 1961 | H$\alpha$ | 0.7 | Telescope | Spectra |
| Nikolskii | 1965 | H$\alpha$ | 0.7 | Coronograph | Spectra |
| Mouradian | 1967 | H$\alpha$ | 1.9 | Telescope | Microphotometer |
| Lynch et al. | 1973 | H$\alpha$ | 0.95 | Telescope | Image |
| Koutchmy & Stellmacher | 1976 | White-Light | 1..2 | total eclipse | Image |
| Ajmanova et al. | 1981 | H$\alpha$ | $\in [0.25, 1.2]$ | partial eclipse | Image |
| Dere et al. | 1983 | EUV | 0.75 | space spectra | Image |
| Cook et al. | 1984 | H$\alpha$ | 1.0 | Telescope | Image |
| Nishikawa | 1988 | H$\alpha$ | $0.61 \pm 0.25$ | Telescope | Filtergram |
| Lorrain & Koutchmy | 1996 | H$\alpha$ | >0..2 | Telescope | Image |
| Budnik et al. | 1998 | N II & N III | 1..8 | space spectra | Spectra |
| Dara et al. ** | 1998 | H$\alpha$ | >0..2 | Telescope | Image |
| Tsiropoula & Tziotziou* | 2004 | H$\alpha$ | 0.72 | Telescope | Image |
| Hansteen et al.* | 2006 | H$\alpha$ | $\in [0.12, 0.72]$ | Telescope | Image |
| De Pontieu et al. **** | 2007 | Ca H | <0.7 | space image | Image |
| Pasachoff et al. | 2009 | H$\alpha$ | $0.66 \pm 0.2$ $\in [0..3,\ 1.1]$ | Telescope | Image |

* Values for dark mottles

** Lower limit of spicule diameter depends of the telescope resolution

## 2- Observation and Image processing of the SOT (Hinode) images.

We use a series of image sequence obtained on 25 October 2008 (near 03h30 U.T.) with the Solar Optical Telescope (SOT) Broadband Filter Imager (BFI) in the Ca II H (centered at 396.86 nm with FWHM of 0.22nm). They have a fixed cadence of 8 seconds (with an exposure time of 0.5 s) and a spatial full resolution corresponding to the SOT of Hinode (Kosugi et al. 2007 & Tsuneta et al. 2008) diffraction limit $\frac{\lambda}{D}$ of $0''.16$ (120km) with a $0.''054$ pixel size.

The size of all frames which were used here is $1024 \times 512$ pixels$^2$ (Hinode read out only the central pixels of the detector to keep the high cadence within the telemetry restrictions) thus covering an area of (FOV) $111'' \times 56''$ at the North Pole. The images are centered at position X=0, Y=948 arcsec, near the pole cap of the Sun where spicule are somewhat more numerous and radial than at low latitudes near the solar equator. They are slightly taller and more systematically oriented nearly the local vertical (Koutchmy and Loucif, 1991; Budnik et al. 1998; Filippov & Koutchmy 2000).

We used the SOT routine program "fg_prep" to reduce the images spikes and jitter effect and aligned time series (Shimizu et al. 2008). The time series show a slow pointing drift, with an average speed less than $0.015''/\mathrm{min}$ toward the north as identified from solar limb motion.

Regarding the image processing, we found superior results after using a spatial image processing for both thread-like (or elongated) and loop like features obtained with the so-called mad-max algorithm (Koutchmy & Koutchmy 1989). The mad-max operator acts to enhance the finest scale structure substantially. The mad-max filter is a weakly nonlinear modification of a second derivative spatial filter. Specifically, it is where the second derivative has a maximum when looking along different directions (usually, 8 directions are used). The behavior of the mad-max qualitatively resembles the second derivative, but the strong selection for the direction of the maximum variation substantially enhances the intensity modulations due the most significant structure and, accordingly, considerably reduces the noisy background like in case of a high spatial filtering. It appears to reduce the blending (due to overlapping effects) between crossing threads superposed along the line of sight. The algorithm, as originally proposed, samples the second derivative in eight directions, but the directional variation of the second derivative was generalized to a smooth function with a selectable passband spatial scale for this work (for more details see November & Koutchmy, 1996). Finally, the choice of the spatial extension of the operator is made taking into account the results shown in table 1 in order not to alter the analyzed structures, as far as an evaluation can be made.

Spatial filtering using "mad-max" algorithms clearly shows relatively bright radial threads in the chromosphere as fine as the resolution limit of about 120 km, see figure 1, as well as some grouping of spicules. Note that some deviation from the radial direction is observed and the aspect ratio of each spicule is often larger than 10.

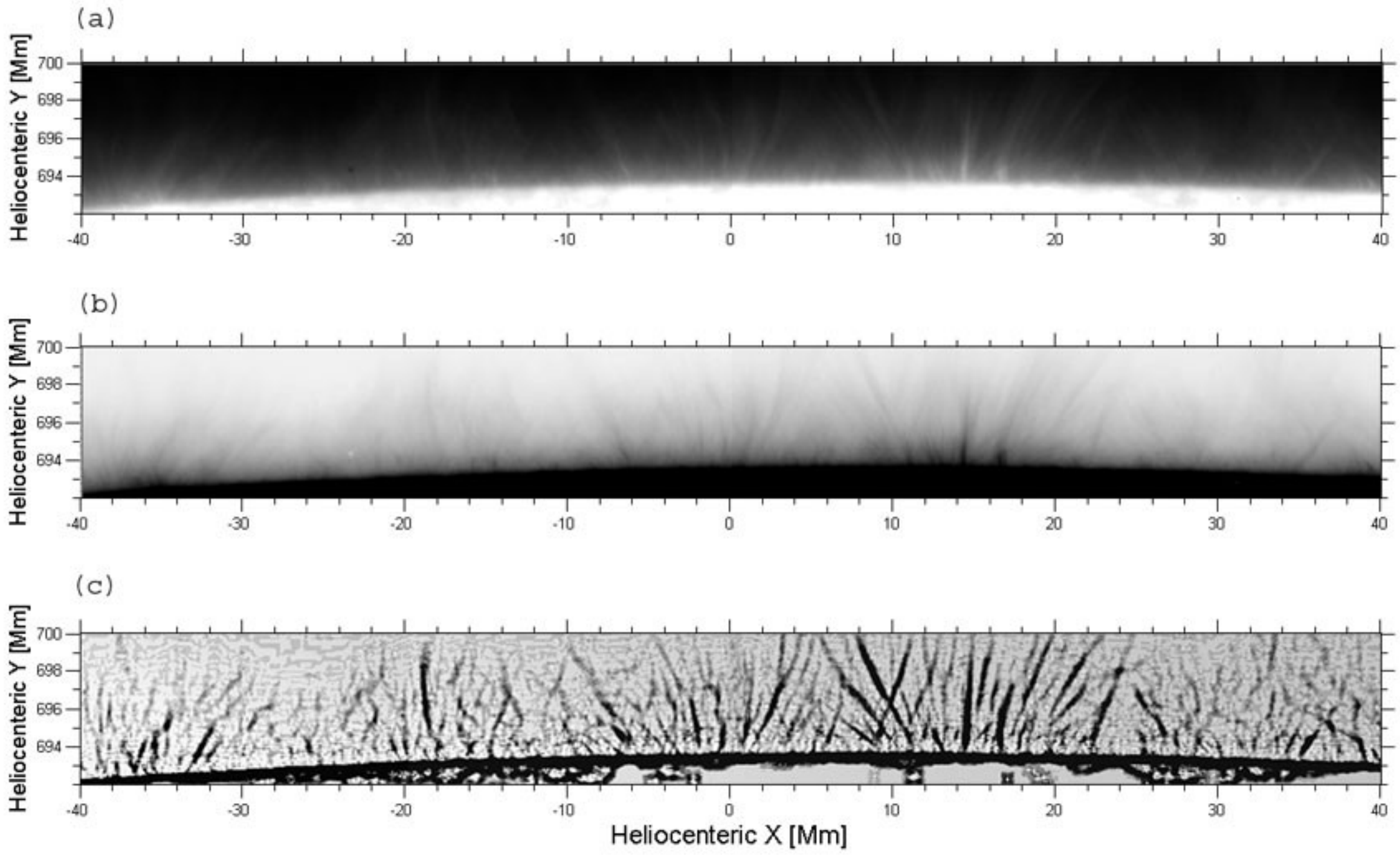

***Fig. 1 a-b-c.*** SOT (Hinode) filtergram in H line of Ca II, obtained on November 25, 2008, near the North pole. (a) is the original images, while (b) is a negative image to improve the visibility of spicules and (c) is the result of applying the mad-max operator.

## 3- Data Reduction & Results

### *3-a* Spicule diameter and spacing.

We computed the Fourier amplitude spectra (AS) of intensity modulations after applying the mad-max operator, see figure 1c, for successive range of heights by step, over the whole time sequence. For each average height, spectra are averaged over 16 pixels, which corresponds to 1″6 arcsec on the image, with a small overlapping between different range of heights, as shown on figure 2. Assuming that the diameter of spicules is constant along their length for each corresponding strictly limited range of heights, doing this averaging reduces the noise without distortion of the resulting spectra but keeping some resolution in height to compare spectra.

In this study, a small spectral smoothing has also been performed by dividing a long series into several segments of equal length of time, computing the power spectra for each segment, and then averaging with the use of an appropriate lag window (Welch 1967).

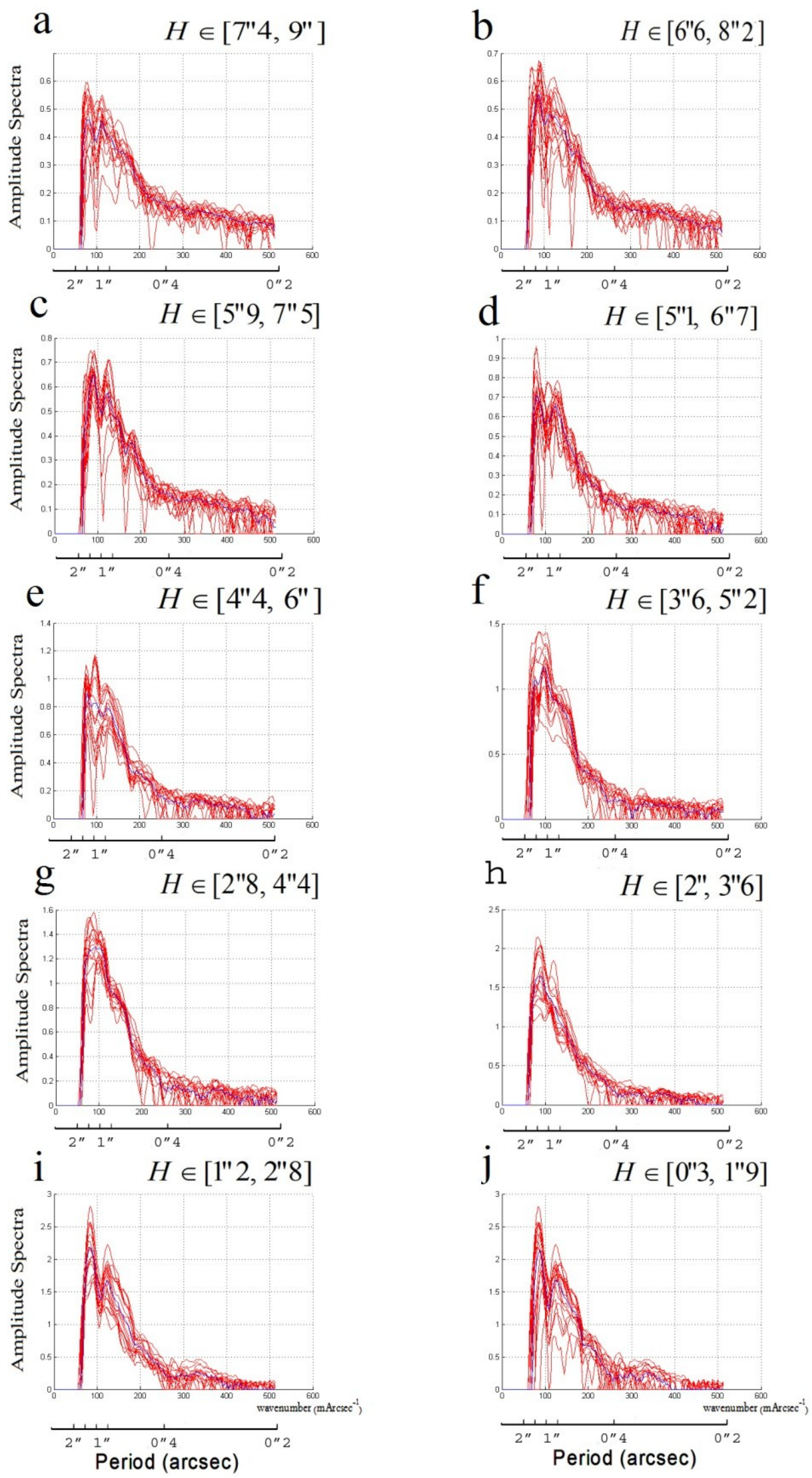

***Fig. 2*** -Amplitude Fourier spectra computed using individual scans of a time series (red lines) with the average in blue lines. The analysis is repeated the same way for different heights above the limb as shown by the range of altitudes H put at the top each graphic.

To have a better visibility of the results, the averaged amplitude spectra deduced for each selected height were plotted together in fig. 3.

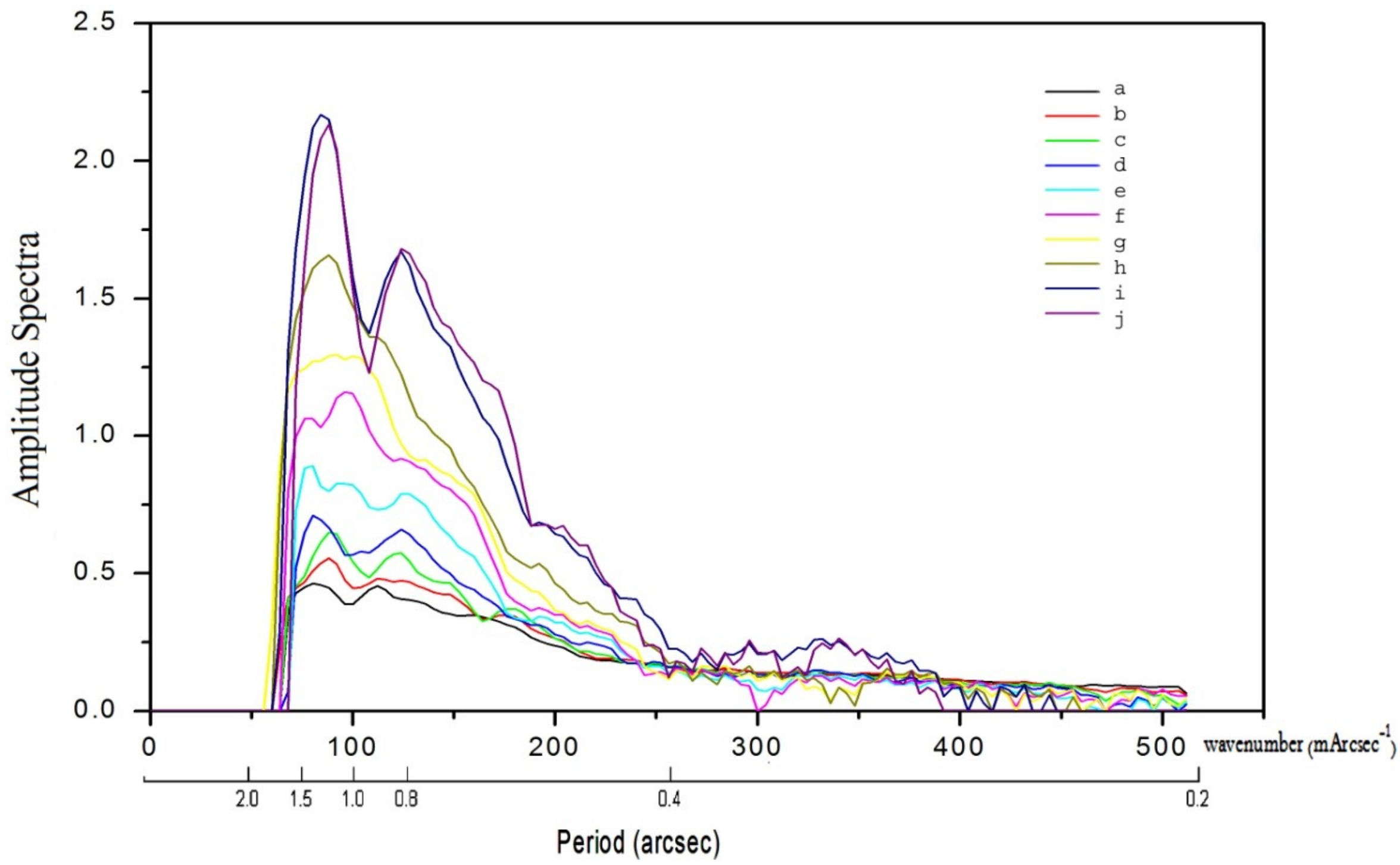

***Fig. 3***- 1D average in time amplitude spectra corresponding to different heights above the solar limb, as shown in fig. 2. Different colors are used for the decreasing heights of fig. 2.

Note that this amplitude Fourier spectra of figure 2 and 3 are not corrected for the modulation transfer function (MTF) of the instrument which attenuates the high frequency part. It is then difficult to quantitatively compare the amplitudes of different spatial frequencies. Qualitatively, it is clear that the amplitudes decrease with the heights with a higher rate near the limb. The true amplitude of the feature in the 0”3 range (0.25 Mm spatial periods or 350 in wavenumber), is indeed considerably higher than what is shown in figure 3 because these short periods are strongly attenuated by the diffraction pattern of the telescope. Correcting the spectra would considerably increase the noise level at high frequencies, without improving the significance level of the results. It could be worth discussing the details appearing on the AS, as far as they are reproducible as shown by the behavior of the AS for different heights. In the lowest levels (i, j), the occurrence of the high frequency part corresponding to spicules of type II near the 0”3 periods (220 km) is striking. Several relative maxima clearly appearing in figure 3 can also be mentioned, near 0”5 (360 km), 0”75 (550 km) and 1”15 (850 km) periods. We believe these features are not just due to statistical fluctuations, because we used a very large amount of spectra to make an average AS and because a rather smooth variation of these features appears with the effective heights. The details of the AS reflect the complex structure of the spicule pattern which could be produced by the behavior of a dynamically unstable flux tube with its skin or wall, where emission is episodically reinforced due to the filling with new plasma or due to a more vigorous heating. It favors the model of a rather empty long flux tube of approximately 800 km diameter with walls of 200 km or less thickness and an azimuthally non uniform distribution. Line of sight effects would give a dispersion in the resulting sizes that are reflected by the AS because the flux tubes- spicules in polar regions are not exactly radial, as seen in figure 1c, see also Koutchmy and Loucif, 1991 and Filippov and Koutchmy, 2000. This 1$^{st}$ approximation speculative model-spicule would naturally explain the occurrence of the doublet structure (in this case the tube would have a wall with an azimuthally uniform density), see Suematsu et al. 2008, and more generally, of interacting spicules that we now will describe in more details.

### ***3-b*** **Multi-component spicules**.

Tanaka (1974) suggested that some spicules which were observed by him on the disk, have two components (in other word he saw doublet threads). Later, high resolution filtergrams obtained by R.B. Dunn at the NSO VTT confirmed this observation, see for ex. Dara et al, 1998. Recently, using the new SOT of Hinode space observations in H CaII taken above the limb, Tanaka's results have been confirmed by Suematsu et al. (2008) which makes really solid this peculiarity of the spicule phenomenon.

High spatial and temporal resolution Hinode images allowed us to now consider the multi-components spicule structures using the method of the time-slice diagrams computed for different heights, see fig. 4-a, b. In the magnified fig. 5 of better visibility, the large blue arrows show the doublet threads. We see that these structures are very close to each other (distance between them is near 1 Mm), the start and the end time for them being almost the same (in other world the life-time is equal) and they are seen in all levels above the limb. All these reasons confirm that they could be two components of the same structure which should be the larger size flux tube.

Furthermore in order to better indentify the waves and measure their propagation period, we followed the time-slice variations (for looking to the time-slice diagram with a better resolution, please use fig. 5). Note that in order to improve the signal/noise ratio in the time slice method, we computed an average intensity along the y-axis direction or height H integrated over a 1 arcsec (~10 pixels) interval.

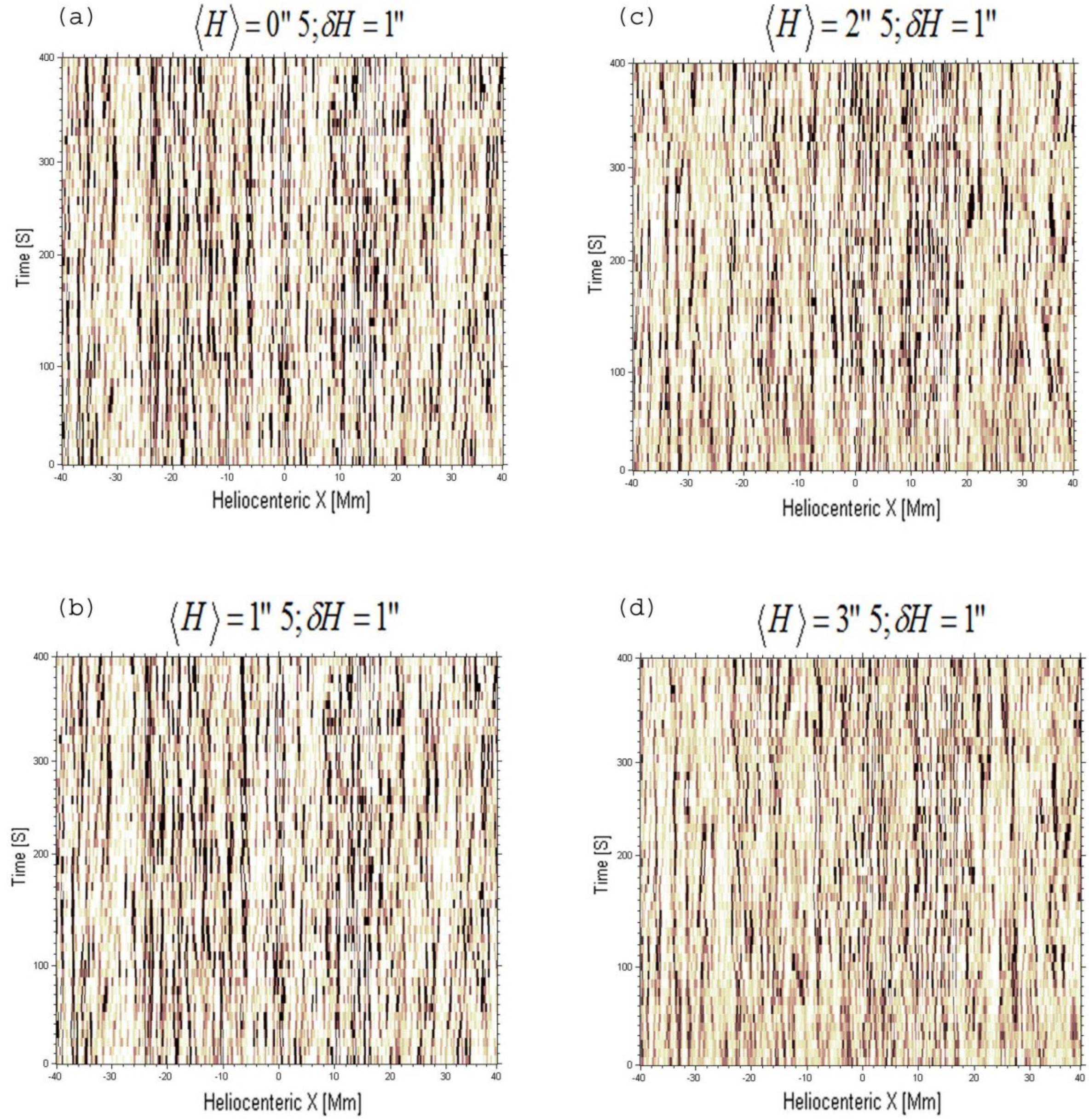

***Fig. 4*-a.** "Time slice" images in the H CaII line, produced using the SOT of Hinode observations closer to the limb. The x axis corresponds to heliocentric coordinates; tick marks in the x-axis correspond to $2''$.

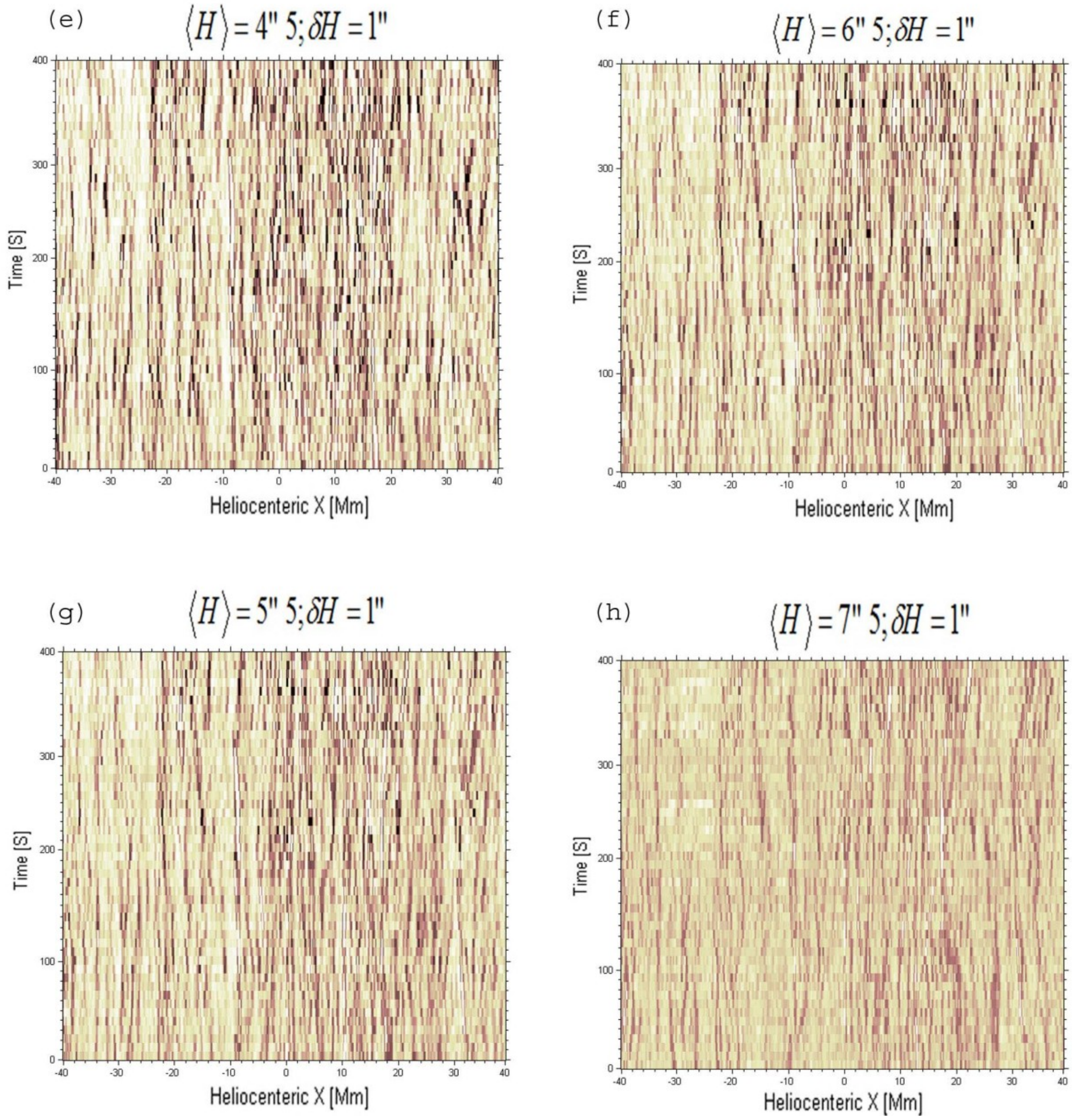

***Fig. 4*-b.** “Time slice” images in Ca H II line, produced using Hinode observations higher up. The x axis corresponds to heliocentric coordinates, tick marks in the x-axis correspond to $2''$.

### *3-c* Alfven waves and helical motions of spicules.

We now recall the spectroscopic observations showing a majority of spectral lines of individual spicules that exhibit inclination or that is tilted with respect to the direction of the dispersion of the spectrograph. Observation of inclined lines in the limb emission spectra of spicules were reported by several authors starting with Michard (1954), Beckers (1966), Noyes (1967) and more extensively by Pasachoff et al. (1968). The spatial resolution of these spectroscopic observations is sometimes low and their interpretation has been controversial, partly because the line of sight integration effect is more important when the smearing is bigger. Using fast spectroscopic limb observations at the NSO R.B. Dunn VTT, Dara et al. (1998) presented some better resolved spectra. It is then interesting to now come back to their interpretation, taking into account our results showing the proper motion of spicules in great details.

The observed tilt of the spectral lines of the spicules can be interpreted as due to spicular rotation. One of the best observations about the spicule rotation has been reported by Noyes (1967). He gave a record of the time development of one of the tilted features observed in the H CaII line at a fixed height above the limb. It is clearly seen that the tilt undergoes a steady evolution of its inclination passing through zero and, afterwards, reversing the direction of inclination. A reasonable explanation of this event must be looked for using a rotational mass motion. In this figure we clearly see that transversal displacements at all heights exist. But we know that in the lower layers of the solar chromosphere the number of spicules is dramatically higher than in the upper layers (Lippincott 1957 & Athay 1959), so the overlapping effect (Ajabshirizadeh et al. 2008) is rapidly decreasing with the height. If the transverse motion was due to the overlapping effect it should have changed in different levels, which is not observed. Similarly, with increasing heights the transverse motion must disappear if it is due to overlapping effect, but from our diagrams we see that the transverse motions in all layers are approximately the same.

The red arrows in fig. 5 was used for showing a possible upward propagating Alfven waves inside a doublet spicule, although the transversal wave fluctuation of the spicule may also be called the fast kink mode oscillation. The kink mode can finally evolve into an Alfven wave after its propagation (Cranmer et al. 2007) and non-linear effects related to the propagation of Alfven waves could be considered. The typical period of the wave is 110 sec.

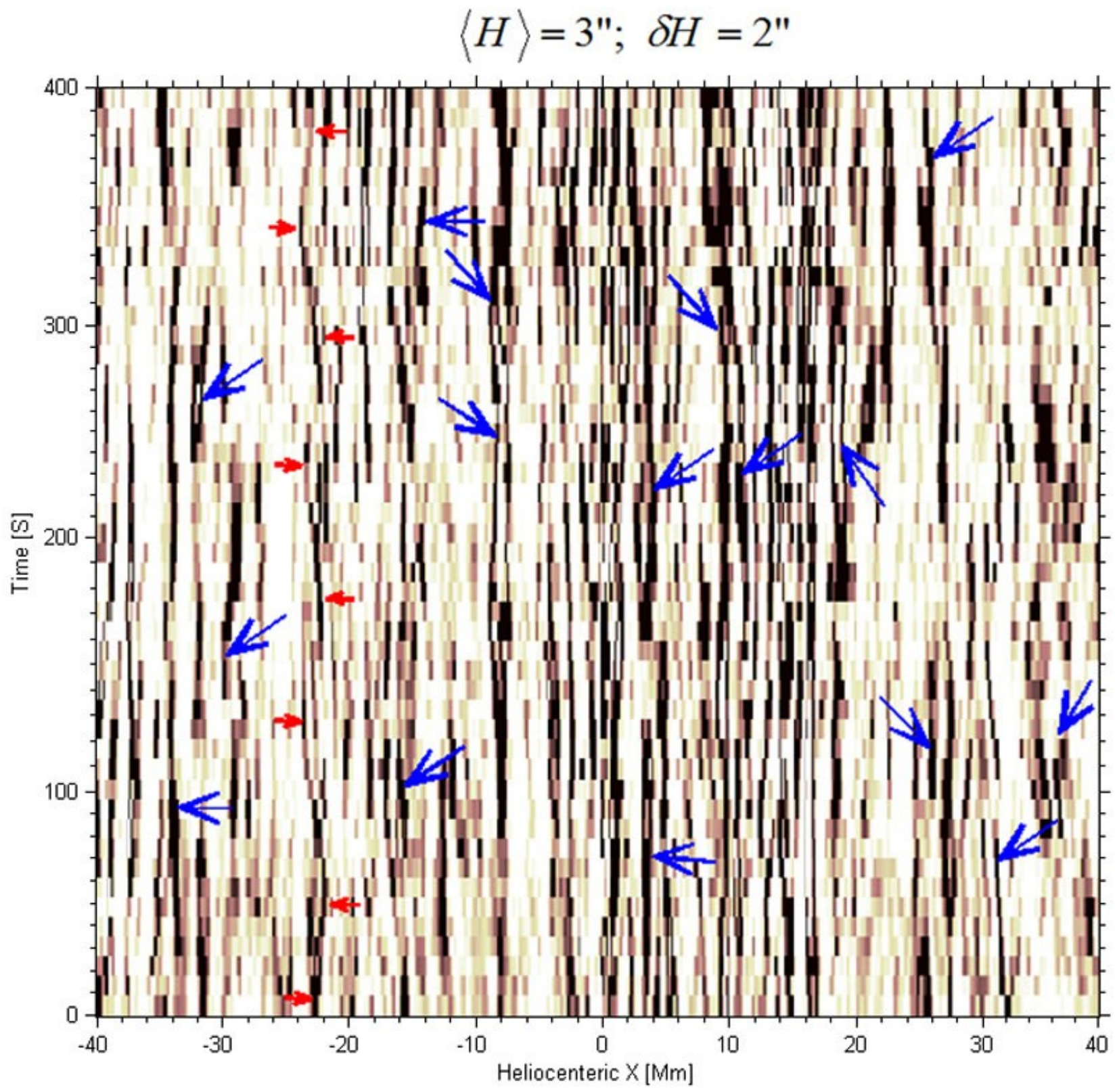

***Fig. 5*** "Time slice" selected image located at the 3 arcsec height above the limb. In order to improve the time slice result and reduce the noise, an average intensity along the y-axis direction taken over a 2 arcsec (20 pixels) interval is shown. The red arrows show a repetitive transverse motion strongly suggesting a twisting or a rotation of a doublet spicule. The doublet threads (spicules) are indicated using blue larger arrows.

Recently, Kukhianidze et al. (2006), De Pontieu et al. (2007b) and He at al. (2009) have reported several different cases of upward propagating high-frequency Alfven waves as identified from the dynamic wave-like behavior of spicules observed by Hinode. The recent rather theoretical paper by Ajabshirizadeh et al. (2009) also shows that the kink waves with periods of ~80-120 s may propagate in spicules at higher level. The occurrence of this range of periods is confirmed by the results presented here which were obtained from time slice diagram for doublet threads. In figure 6 we clearly see the lateral motion with a period of 110 s of a typical doublet threads that we selected for illustration.

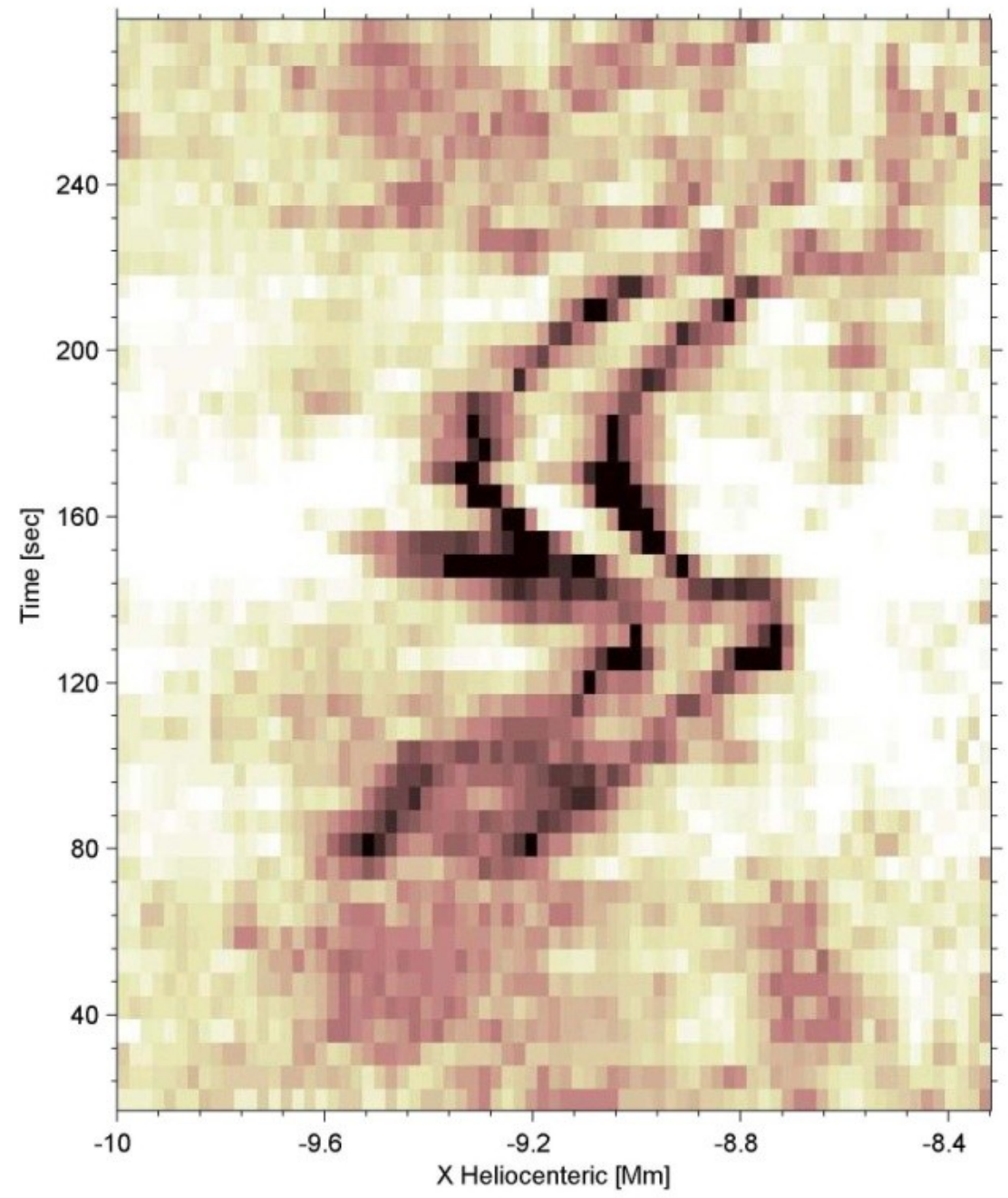

***Fig. 6-*** "Time slice" image in negative from the HCa II line SOT observations from Hinode. The x axis corresponds to heliocentric coordinates in Mm. The wave like pattern of a doublet spicule is dramatically illustrated.

We know that the torsional Alfven mode do not alter the geometry of the loop, i.e., no displacement of the loop axis or deformations of the loop width (Doorsselaere et al. 2008). However, the global Alfven waves (of the whole flux tube), which may indeed fill a significant volume in and around each spicule, may cause the global transverse oscillation of the magnetic field lines.

We should however be careful about the interpretation of time slice diagram results. It is important that we separate the rotational structures from the transversal motion phenomena; because these two phenomena could show very similar effects in the time slice diagram. In our case, after comparing what is shown on the time slice of figure 5 with red arrows and coming back to the Hinode image movie, we found that the effect is very probably due to the spinning or twisting motion of a typical doublet thread- spicule. This spin motion of doublet threads was indeed already reported by Suematsu et al. (2008). This helical motion of spicules needs in more considerations in future works before a definite model, taking into account the values of the diameter and the thickness of the wall of the flux tube can be proposed.

## 4. Discussion and conclusions

In order to determine the true diameter of solar spicules we used Hinode image to perform some more statistically significant measurements. According to the results of the Fourier analysis, the width of the spicules ranges from ~200 km to more than 1000 km, which in agreement with both the historical results (see table I) and the most recent reports of De Pontieu et al (2007).

The reduction of the much improved spicule filtergrams from SOT onboard Hinode leads to the diameters that are not substantially different from values found before, taking into account the multiple structure of spicules in grouping and in doublet structure. In figure 2 and more clearly in figure 3, we see that the amplitude Fourier spectra show a significant peak near 200 km for the nearest to the limb layers. They should correspond to the type II spicules very recently discovered by De Pontieu et al. (2007). Other peaks of amplitude power were seen for all heights of the atmosphere, their values showing a typical diameter for spicule near 1 arcsec. Indeed spicules show a whole range of diameters, including "interacting" spicules (I-S), depending of the definition chosen to characterize this ubiquitous dynamical and magnetic phenomenon. From our plots different values have been obtained for what can be called the spicule diameter; in our opinion all of them are correct and they correspond to the type I and type II spicules introduced by De Pontieu et al. (2007 a & b). However from the discussion of these results a model of spicule- flux tube seems to emerge.

We have also found in several cases the doublet threads that Tanaka (1974) reported from on disk observations a long time ago and more recently reported by Suematsu et al. (2008) where several off limb doublet threads using Hinode observation were shown.

From time slice diagram we see several doublet threads. We speculate on these doublet structures as being the two components of the same flux tube; then in general spicules have a multi-component structure, so doublet threads could belong to the single wide magnetic flux tube with walls. This would permit a more solid approach to the problem of their dynamics, including the question of waves propagating along spicules, and their oscillatory behavior.

Fig. 4 and 5 show a lot of doublet threads occurring in all heights and we know that the overlapping effect is not important for high levels (Ajabshirizadeh et al. 2008).

Accordingly, based on our analysis after applying the mad-max operator to all the SOT-Hinode spicule images, the qualitative impression suggests that threads are approximately cylindrically symmetric in three dimensions flux tubes. This is not in contradiction with the classical theoretical models similar to what was described for ex. in Lorrain & Koutchmy (1996) although today the physics at the feet of spicules could more include in more details the process of emergence of very small magnetic loops with a null point above. Some spicules show high-frequency lateral motion, and we report evidence for a rotational behavior of some spicule as known from the spectroscopic data where inclined spectral lines in the limb emission spectra of spicules were reported due to mass rotation effect of the spicule- flux tube. Time-slice plots of Ca II H time series showed several impressive cases of rotational multi-component structures as again illustrated in figure 7.

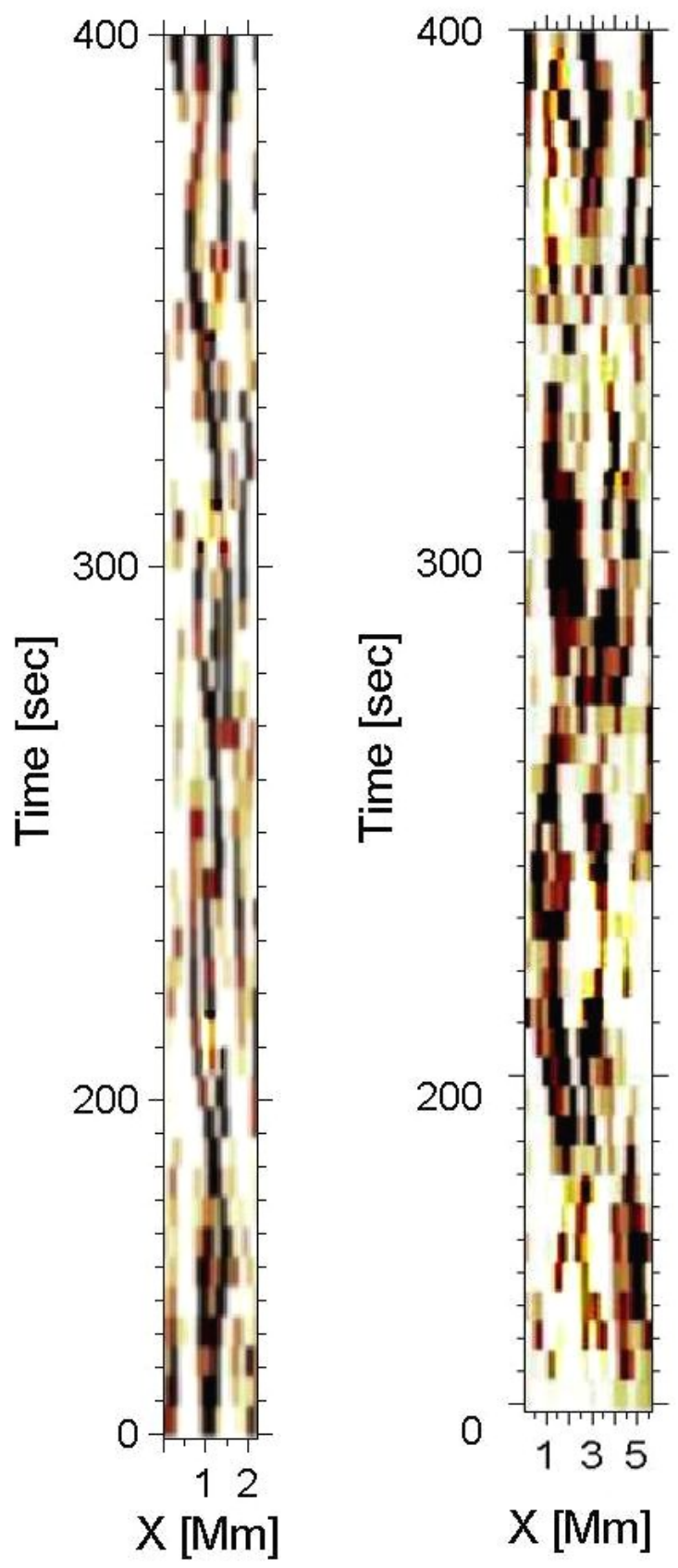

***Fig. 7-*** "Time slice" plots which are located at 4 arcsec height above the limb. In order to improve the signal/noise level, we used the average intensities along the y-axis direction taken over 1 arcsec (10 pixels) interval, which introduces a negligible smearing in the results of the analysis.

Moreover, most observers of spicule spectra noticed a large number of cases in which the spicule emission line lies not perpendicular to the spectrograph dispersion, but make a slight angle with it (Michard 1954, Nikolsky, 1965, Livshitz 1966, Beckers 1968, Pasachoff et al. 1968 and Avery 1970). Such observations of a tilt of the emission line have been interpreted as resulting from the rotation of a rather wide spicule. Pasachoff et al. (1968) estimate the spin velocity of spicule components to be about 30 km/sec.

Suematsu et al. (2008) have also reported the observation of a twist motion of spicules from Hinode SOT image observation and He et al. (2009) reported high frequency transversal motion in spicule, in agreement with our more extended results shown on figures 4 to 7. In fig. 5 for ex. we indicated this behavior using red arrows and showed (right side, at heliocentric $X \in [35,40]$ Mm) a special structure that is more typical of a helical motion. Then, the SOT Hinode observations, giving an unprecedented spatio- temporal resolution on images taken above the limb, brought, after a suitable processing of images is made, a new insight in the difficult problem of the occurrence of dynamical cool spicules as an important component of the ubiquitous flux tubes permeating the whole extended chromosphere. It is a set of rich material that will certainly be of use for the theoretical modeling of this mysterious region situated at the interface between the surface and the very hot and extended corona.

Finally, this new view on spicules is important when discussing the large range of spicule diameters obtained using the historical ground-based telescope observations with lower resolution due to the earth atmosphere seeing conditions. It is then quite difficult to detect the fine components separately and they will be seen as a single structure, provided some coherence exists between the different components. This is the explanation which comes in minds to explain all the historical observations made when the details of the spicule- flux tube were not resolved. Using fast imaging and/or adaptive optics systems the difficulty can partly be overcome and some new spectroscopic observation will hopefully complement the imaging observations reported here to provide a more definite diagnostics.

**Acknowledgements.** We thank L. Ofman, R. Erdelyi, L. Golub and J. Pasachoff for useful discussions and we are very grateful to the Hinode team for providing wonderful observations. Hinode is a Japanese mission developed and launched by ISAS/JAXA, with NAOJ as domestic partner and NASA and STFC (UK) as international partners. Image processing Mad-Max program was provided by O. Koutchmy, see http://www.ann.jussieu.fr/~koutchmy/index_newE.html. This work has been supported by RIAAM.